\documentclass{article}
\usepackage{graphicx} 
\usepackage{subcaption}
\usepackage{float}
\usepackage{geometry}
\usepackage{amsmath}
\usepackage{amssymb}
\usepackage{amsfonts}
\usepackage{physics}
\usepackage{xcolor}
\usepackage{appendix}
\usepackage{hyperref}
\hypersetup{
    colorlinks = True,
    linkcolor = blue,
    filecolor = magenta,
    urlcolor = cyan,
    pdftitle = {Overleaf Example}, 
    pdfpagemode = FullScreen
}
\usepackage{comment}
\usepackage{authblk}
\usepackage{caption}
\begin{document}

\title{(Gravitational Wave) Memory of Starobinsky in a Time Crystal (Condensate)}

\author[1]{Aurindam Mondal\thanks{\texttt{aurindammondal99@gmail.com}}}
\author[2]{Subir Ghosh\thanks{\texttt{subirghosh20@gmail.com}}} 
\affil{\makebox[\textwidth][c]{\small{$^{1}$Physics and Applied Mathematics Unit, Indian Statistical Institute,}} \newline 
          \makebox[\textwidth][c]{\small{203, Barrackpore Trunk Road, Kolkata, India-         700108.}}}
          

\date{\vspace{-1.4 cm}}

\maketitle

\abstract{In this Letter we have revealed  the presence of Gravitational Wave Memory Effect (GWME) in a new and physically appealing scenario  - the Time Crystal (TC) condensate of Starobinsky Quadratic Gravity. We have used a Gravitational Wave form, induced in a TC condensate in FLRW spacetime, that is much more general than the idealized and somewhat unphysical Plane Wave spacetimes, that are conventionally used. We have presented new results for non-zero  GWME, both  in coordinate and in velocity variables. The results are expressed in Cartesian and Brinkmann coordinates. Approximate analytic forms of transverse geodesics in Brinkmann coordinates are  provided. Very rough quantitative estimates for GWME are also suggested.}

\vspace{0.6 cm}

{\bf{Introduction:}} Existence of Gravitational Waves (GW) was predicted by Einstein himself \cite{ein} but its first  direct observation  was made only about 10 years ago by  the LIGO and Virgo collaborations \cite{gw}. Obviously the principal reason for this delay is the smallness of the signature: the gravitational wave signal  GW150914, generated from the inward spiral and merger of two black holes, $\sim 36\times $ and $29\times$ solar mass and the subsequent ringdown of a single, $62\times$ solar mass remnant  black hole,  created a ripple   in spacetime on earth, that managed to change the length of a 4 km LIGO arm  by $10^{-18}m \sim $ a thousandth of the width of a proton! 

Gravitational Wave Memory Effect (GWME), a persistent change in the position and velocity of a set of particles (or detectors), due to the passing of a GW, was predicted in \cite{zel} in linearized general relativity (although see also \cite{new}), and it was extended  beyond the weak-field approximation, to a non-linear memory effect in  \cite{chris}. These effects were unified in \cite{fav}. Till date there is no smoking gun evidence in favour of GMWE although  most promising setups are Pulsar Timing Arrays, LISA, and ground-based interferometers such as      LIGO/Virgo/KAGRA.

Theoretical predictions in somewhat idealized  examples have appeared  \cite{kar} following the geodesic separation analysis \cite{gib}. Unfortunately most of these suffer from a drawback: for computational simplicity, these works take recourse to {\it{Plane Wave}} spacetimes, which are analogous to  scalar or electromagnetic plane waves  in flat spacetime. However, as is well known, from  a global point of view, these GW spacetime models  are not  physically reasonable, as they are not asymptotically flat, or not even globally hyperbolic,  carries an infinite amount of energy, and are not related to sources. Hence,  the nice analytic expressions in these artificial spacetimes can, at best, claim the consistency of GWME in General Relativity. At this crossroad, our aim is to provide concrete evidence of GWME in a (not so clean, but) realistic and viable GW spacetime, notably the Time Crystal  Condensate (TC) scenario in Starobinsky $R^2$-model, studied earlier by us \cite{aur}. 

To recapitulate briefly, a new phase of (not fully in thermodynamic equilibrium)  matter - Time Crystal (TC) - was proposed by Wilczek \cite{wil} in quantum system and by Shapere and Wilczek \cite{wil1} in classical system, (relevant to the present work), that can have non-trivial dynamics in its (quantum) ground state or (classical) lowest energy state. Closely followed an alternative formulation of TC \cite{sg} where the dynamical lowest energy state was induced via a Spontaneous Symmetry Breaking in momentum space, generated in a {\it{higher derivative}} theory. (See \cite{sach} for an exhaustive review on TC). The Starobinsky cosmological model \cite{star}, in a weak gravity limit, decouples into a spin $2$ graviton and a higher derivative scalar field \cite{alv}. In \cite{sg}, a scenario was considered where this scalar acted as a condensate background where the graviton lives. Furthermore, it was emphasized that, apart from the metric, no other matter degrees of freedom was introduced and  the  higher derivative nature of the Starobinsky model was enough to generate this dynamical condensate. The FLRW cosmology in the presence of this TC Condensate was studied in \cite{scsg} where an explicit form of the scale factor  $a(t)$ was derived, depending on the $R^2$-coupling parameter. Subsequently, GW form in this TC condensate was studied in \cite{aur}. (For a recent perspective on Dark Energy - Time Crystal connection see \cite{laura}.)

Highlights of the present Letter are
\begin{itemize}
\item  Presence of GWME  revealed in a new arena - the TC condensate of Starobinsky Quadratic Gravity.
\item The spacetime considered is qualitatively different and much more general than the Plane Wave spacetime conventionally used.
\item  GWME, both  in coordinate and in velocity, are expressed  in Cartesian and Brinkmann coordinates. Approximate analytic forms of transverse geodesics in Brinkmann coordinates are  provided.
\item Very rough quantitative estimates for GWME are  suggested.
\end{itemize} 

\vspace{0.1 cm}

{\bf{Geodesics in TC Condensate FLRW background:}} Instead of directly solving the (second order) geodesic equation, it is advisable to exploit the  symmetries of the given  spacetime to generate first order equations of motion. Lagrangian of a freely falling particle (of mass scaled to unity) for an FLRW universe in presence of TC condensate,  
\begin{equation}\label{Vacuum_Lagrangian}
    \mathcal{L} \, = \, g_{\mu\nu} \, \frac{dx^{\mu}}{d\tau} \frac{dx^{\nu}}{d\tau}
    \hspace{0.3 cm} = \hspace{0.3 cm} - \left(\frac{dt}{d\tau} \right)^{2} \, + \, a^{2}(t) \left\{\left(\frac{dx}{d\tau} \right)^{2} + \left(\frac{dy}{d\tau} \right)^{2} + \left(\frac{dz}{d\tau} \right)^{2} \right\} \, . 
\end{equation}
It enjoys a set of cyclic coordinates $x, y$ and $z$, with the conserved conjugate momenta $p_i=\partial \mathcal{L}/\partial \dot{x}_{i}=2a^2~dx_i/d\tau =p_{i0}$, where $a(t)$ is the scale factor and $\tau$ is the proper time. Proper time $\tau$ can be replaced by cosmic time $t$ using  the normalisation of the relativistic four velocity
\begin{equation}
     g_{\mu\nu} \, \frac{dx^{\mu}}{d\tau} \frac{dx^{\nu}}{d\tau} \hspace{0.2 cm} = \hspace{0.2 cm} -1 \hspace{0.4 cm} \Rightarrow \hspace{0.4 cm} 
     \frac{dt}{d\tau} \hspace{0.2 cm} = \hspace{0.2 cm} \sqrt{1 + a^{2} \left\{\left(\frac{dx}{d\tau} \right)^{2} + \left(\frac{dy}{d\tau} \right)^{2} + \left(\frac{dz}{d\tau} \right)^{2} \right\}} \, ,  
\end{equation}
which leads to the following first order geodesic equations, 
\begin{eqnarray}
    \frac{dx_i}{dt} \, = \, \frac{dx_i}{d\tau} \, \frac{d\tau}{dt} \hspace{0.2 cm} = \hspace{0.2 cm} \frac{(p_i)_0}{2 a^{2}} \hspace{0.1cm} \frac{1}{\sqrt{1 + \left\{(p_x)^{2}_{0} + (p_y)^{2}_{0} + (p_z)^{2}_{0} \right\}/4 a^{2}}} \,  
    \label{ee}
\end{eqnarray}
where $(p_i)_0 =p_i(t=t_0)$. Eq (\ref{ee}) can be solved to get the coordinates $x_{i}(t)$ of freely falling particle for an explicit  form of scale factor $a(t) \approx a(t_{0}) \sqrt{1 + 2 H_{0}(t-t_{0})}$ (see Appendix A), derived for FLRW universe in presence of TC condensate \cite{aur}. \\ 

{\bf{ GW profile in TC Condensate FLRW background :}} The GW profile  propagating in a TC Condensate FLRW background has already been studied in \cite{aur} (see Appendix), as follows: up-to first order perturbation (about the TC Condensate background),  the Einstein  equations yield
\begin{eqnarray}\label{GW_equation}
    \Box\tilde{h}_{\alpha\beta} - 2\overline{R}_{\nu\beta\alpha\mu} \tilde{h}^{\nu\mu} - \overline{R}_{\nu\alpha} \tilde{h}^{\nu}_{\beta} - \overline{R}_{\nu\beta} \tilde{h}^{\nu}_{\alpha} &=& 0 \, . 
\end{eqnarray}
Here $\tilde{h}_{\alpha\beta}$ represents the trace reversed  part of metric perturbation, defined as $\tilde{h}_{\alpha\beta} = h_{\alpha\beta} - (1/2)\overline{g}_{\alpha\beta} h$ and $\overline{R}_{\alpha\beta\mu\nu}$ represents the background Riemann tensor with the background metric $\overline{g}_{\alpha\beta}$. In Traceless-Transverse gauge, the  line element  in presence of GW profile is parametrised by
\begin{eqnarray}\label{metric_GW}
    ds^{2} &=& - \, dt^{2} + a(t)^{2} \, dx^{2} + \left\{a(t)^2 + h_0(t,x) \right\} \, dy^{2} + \left\{a(t)^2 - h_0(t,x) \right\} \, dz^{2} + 2h_{1}(t, x) \, dydz \, . 
\end{eqnarray}
where  $h_{0}(t,x)$ and $h_{1}(t,x)$ are explicitly given by \cite{aur}
\begin{eqnarray}
    h(t,x) \, = \, h_{0}(t,x) &=& h_{1}(t,x) \hspace{0.2 cm} \approx \hspace{0.2 cm} c_{3} c_{2} \hspace{0.1cm} e^{-A+A\ln{(A)}} \bigg(\frac{2 \sqrt{41\Omega_{\lambda}}}{A^{1/3}} \bigg)^{3/4} \hspace{0.1cm} \frac{e^{-i\sqrt{\Gamma}x}}{\sqrt{4\Delta t}} \hspace{0.1cm} \sin{\bigg\{2\sqrt{A\Delta t} - \big(4A+1 \big)\frac{\pi}{4} \bigg\}} \hspace{0.08cm} .
    \label{33}
\end{eqnarray}
Here $c_{3}, c_{2}$ are the integration constants of perturbed Einstein field equations and $\Omega_{\lambda}, \Omega_{k}, \Omega_{r}$ are the density parameters corresponding to effective cosmological constant, spatial curvature and radiation like terms. On the other hand, $A$ and $\Delta t$ are defined as, 
\begin{eqnarray}
    A \, = \, \frac{4 \hspace{0.06cm} \Gamma - 27a^{2}_{0}H^{2}_{0} \hspace{0.06cm} \Omega_k}{8 a^{2}_{0} H^{2}_{0} \sqrt{41\Omega_{\lambda}}}  \hspace{0.4 cm} ,~~ \hspace{0.2 cm}
    \Delta t \, = \, \frac{\sqrt{41\Omega_{\lambda}}}{2} \left\{1 + 2H_{0}(t-t_{0}) \right\} \hspace{0.06cm} . 
\end{eqnarray}
Here $\Gamma$ be the separation constant arising from the separation of variables method of gravitational wave equation Eq (\ref{GW_equation}) and $a_{0}, H_{0}, t_{0}$ are the present day value of scale factor, Hubble parameter and age of the universe respectively (Some details are provided in Appendix A). 

Transforming the above mentioned perturbed FLRW metric Eq (\ref{metric_GW}), written in the coordinates $\{t, x, y, z\}$, into a diagonal one, written in another coordinates $\{t_{1}, x_{1}, y_{1}, z_{1} \}$ (see Appendix B), Lagrangian of the same  freely falling particle in TC Condensate FLRW becomes, 
\begin{eqnarray}\label{diagonal_Lagrangian}
    \mathcal{L} &=& - \left(\frac{dt}{d\tau} \right)^{2} + a^{2} \left(\frac{dx}{d\tau} \right)^{2} + \lambda_1 \left(\frac{dy}{d\tau} \right)^{2} + \lambda_2 \left(\frac{dz}{d\tau} \right)^{2} \hspace{0.08cm} , \nonumber \\ 
    \lambda_1(t,x) &=& a(t)^2+\sqrt{2} h(t,x) \hspace{0.7 cm} , \hspace{0.7 cm} \lambda_2(t,x) \hspace{0.2 cm} = \hspace{0.2 cm} a(t)^2-\sqrt{2} h(t,x) \, . 
\end{eqnarray}
Here the transformed coordinates $\{t_{1}, x_{1}, y_{1}, z_{1} \}$ are again renamed as $\{t, x, y, z \}$. \\ 

{\bf{Geodesics in  presence of GW spacetime (Cartesian coordinate):}} The geodesic equation is 
\begin{equation}
    \frac{d^{2}x^{\mu}}{d\tau^{2}} \, + \, \Gamma^{\mu}_{\alpha\beta} \, \frac{dx^{\alpha}}{d\tau} \frac{dx^{\beta}}{d\tau} \hspace{0.2 cm} = \hspace{0.2 cm} 0 . 
    \label{ge}
\end{equation}
However, in presence of GW profile Eq (\ref{diagonal_Lagrangian}),  $y$ and $z$ still remain cyclic coordinates with the conserved momenta  $(p_y)_0$ and $(p_z)_0$ respectively, leading to the following first order dynamical equations
\begin{eqnarray}\label{geod_yz}
    \frac{dy}{d\tau} \hspace{0.2 cm} = \hspace{0.2 cm} \frac{(p_y)_0}{2\lambda_1} \hspace{1.2 cm} , ~~ \hspace{1.4 cm} 
    \frac{dz}{d\tau} \hspace{0.2 cm} = \hspace{0.2 cm} \frac{(p_z)_0}{2\lambda_2} \hspace{0.08cm} . 
\end{eqnarray}
Due to the $x$-dependent  GW profile, there is no conserved momentum conjugate to the x-coordinate. Utilizing the Eq (\ref{ge}), the x-geodesic is found to be
\begin{eqnarray}\label{geod_x}
        2 a^{2} \frac{d^{2}x}{d\tau^{2}} + 4a a' \left(\frac{dt}{d\tau} \frac{dx}{d\tau} \right) + \sqrt{2} \, \frac{\partial h_{0}}{\partial x} \left\{\left(\frac{dz}{d\tau} \right)^2-\left(\frac{dy}{d\tau} \right)^2 \right\} = 0 \hspace{0.08 cm} . 
\end{eqnarray}
Similarly, the t-geodesic is given by  
\begin{eqnarray}\label{2nd_order_t}
        2 \, \frac{d^{2}t}{d\tau^{2}} + 2a a' \left\{\left(\frac{dx}{d\tau} \right)^{2} + \left(\frac{dy}{d\tau} \right)^{2} + \left(\frac{dz}{d\tau} \right)^{2} \right\} + \sqrt{2} \, \frac{\partial h_{0}}{\partial t} \left\{\left(\frac{dy}{d\tau} \right)^{2} - \left(\frac{dz}{d\tau} \right)^{2} \right\} &=& 0 \hspace{0.08 cm} . 
\end{eqnarray}
We perform the same normalisation exercise as before, 
\begin{eqnarray}\label{Normalisation_momenta}
     g_{\mu\nu} \, \frac{dx^{\mu}}{d\tau} \frac{dx^{\nu}}{d\tau} \, = \, -1 \hspace{0.6 cm} \Rightarrow \hspace{0.6 cm}
   \frac{dt}{d\tau} \, = \, \pm \hspace{0.08cm} \sqrt{\frac{1 + \lambda_1\left(dy/d\tau \right)^{2} + \lambda_2\left(dz/d\tau \right)^{2}}{1 - a^{2} (dx/dt)^{2}}} .   
\end{eqnarray}
Substituting the proper time $\tau$ by the cosmic time $t$ through the Eq. (\ref{2nd_order_t}) and (\ref{Normalisation_momenta}), all the geodesic equations in cosmic time $t$ are rewritten as follows
\begin{equation}
\frac{dy}{dt}=\frac{Q}{P\lambda_1}(p_y)_0 \hspace{0.7 cm} , \hspace{0.7 cm} \frac{dz}{dt}=\frac{Q}{P\lambda_2}(p_z)_0 \, , 
\label{rer}
\end{equation}
\begin{equation}
    \left[2a^{2} \, \frac{d^{2}x}{dt^{2}} + 4a\dot a \, \frac{dx}{dt} - 4a^{3}\dot a \ \left(\frac{dx}{dt} \right)^{3} \right] \hspace{0.1cm} \boldsymbol{=} \frac{Q^2}{P^2}
      \left[\sqrt{2} \left(\frac{\partial h}{\partial x} + 2a^{2} \frac{dx}{dt} \frac{\partial h}{\partial t} \right) \left\{\frac{(p_y)_0^{2}}{\lambda_1^{2}} - \frac{(p_z)_0^{2}}{\lambda_2^{2}} \right\} + 4a^{3}\dot a \frac{dx}{dt} \left\{\frac{(p_y)_0^{2}}{\lambda_1^{2}} + \frac{(p_z)_0^{2}}{\lambda_2^{2}} \right\} \right] . \hspace{0.8cm} 
    \label{22}
\end{equation}
where $P=\left\{4 + \frac{(p_y)_0^{2}}{\lambda_1} + \frac{(p_z)_0^{2}}{\lambda_2} \right\}^{1/2} \, , \, Q= \left\{1 - a^{2} \left(\frac{dx}{dt} \right)^{2} \right\}^{1/2}$. Neglecting the non-linear terms of the x-geodesic and considering the metric perturbations to be small enough, one gets the following simplified form of x-geodesic, with 
$\psi_{0} = \left(\tilde{p}_{z0}^{2} - \tilde{p}_{y0}^{2} \right)$ and $\psi_{1} = \left(\tilde{p}_{z0}^{2} + \tilde{p}_{y0}^{2} \right)$, 
\begin{eqnarray}
    \left(2a^{2} \, \frac{d^{2}x}{dt^{2}} + 4a\dot a \, \frac{dx}{dt} \right) \left[1 + \frac{1}{4a^{2}} \left(\psi_{1} + \frac{\sqrt{2}\psi_{0} h}{a^{2}} \right) \right] &=& \frac{\dot a}{a} \frac{dx}{dt} \left(\psi_{1} + \frac{2\sqrt{2}\psi_{0} h}{a^{2}} \right) - \frac{\psi_{0}}{2\sqrt{2}a^{4}} \left(\frac{\partial h}{\partial x} + 2a^{2} \frac{\partial h}{\partial t} \frac{dx}{dt} \right) \, . \hspace{0.8cm}
\end{eqnarray}

\begin{figure}[H]
    \centering
    \includegraphics[width=0.32\linewidth]{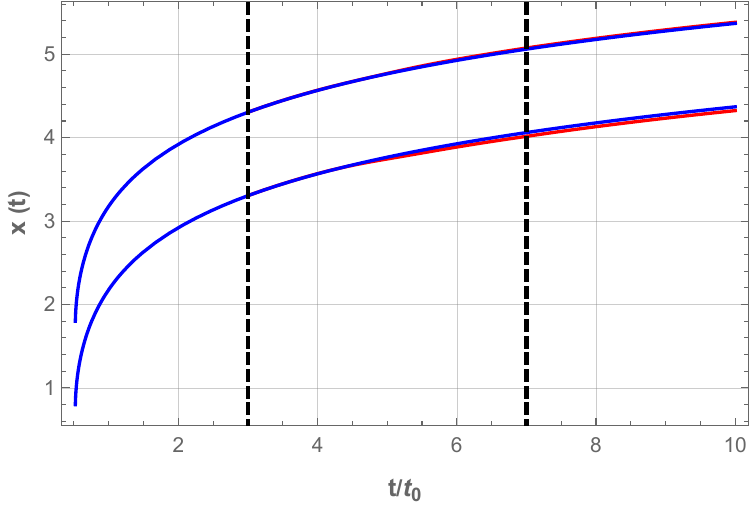}
    \includegraphics[width=0.32\linewidth]{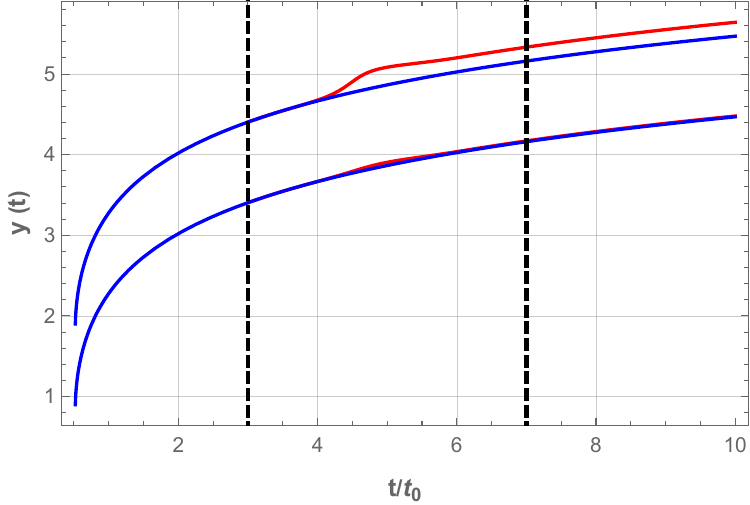}
    \includegraphics[width=0.32\linewidth]{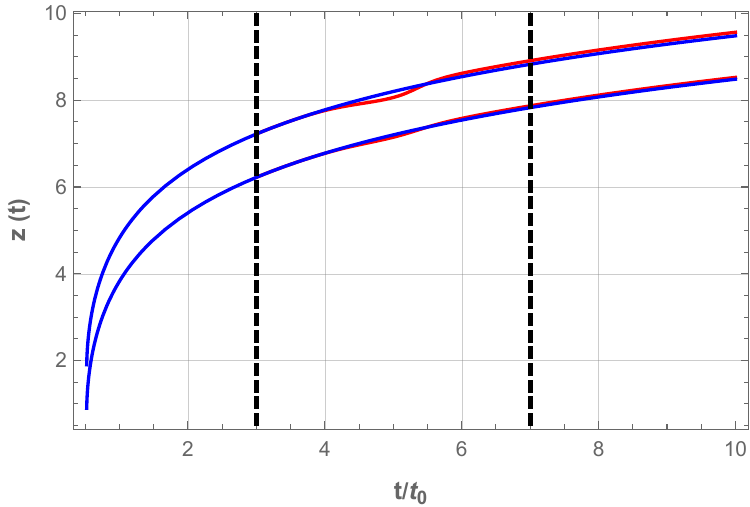}
    \caption{The x, y and z-coordinate of geodesic equation are plotted as a function of cosmic time $t/t_{0}$. Blue curve denotes the results in absence of GW profile and Red curve denotes the same in  presence of GW profile.} 
    \label{fig:1}
\end{figure}

\begin{figure}[H]
    \centering
    \includegraphics[width=0.32\linewidth]{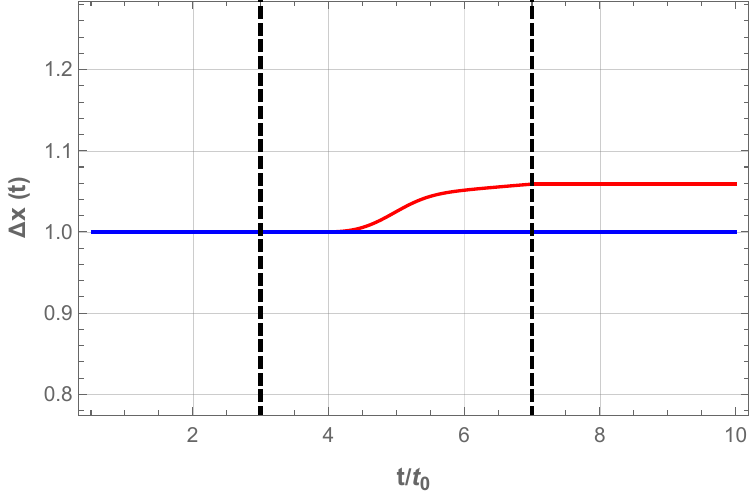}
    \includegraphics[width=0.32\linewidth]{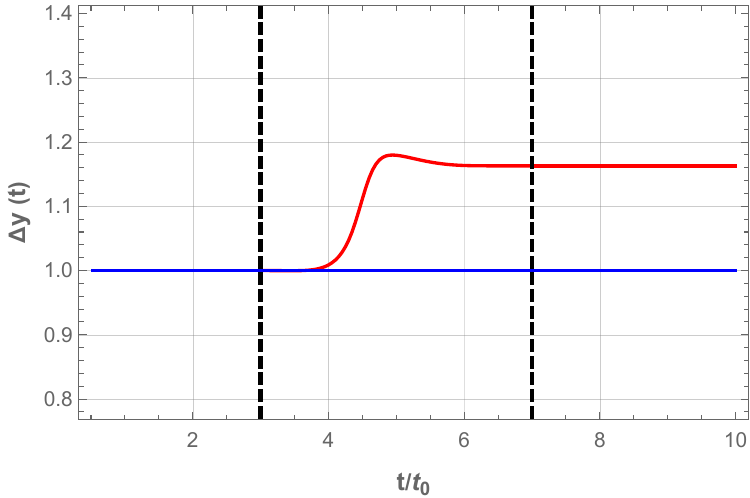}
    \includegraphics[width=0.32\linewidth]{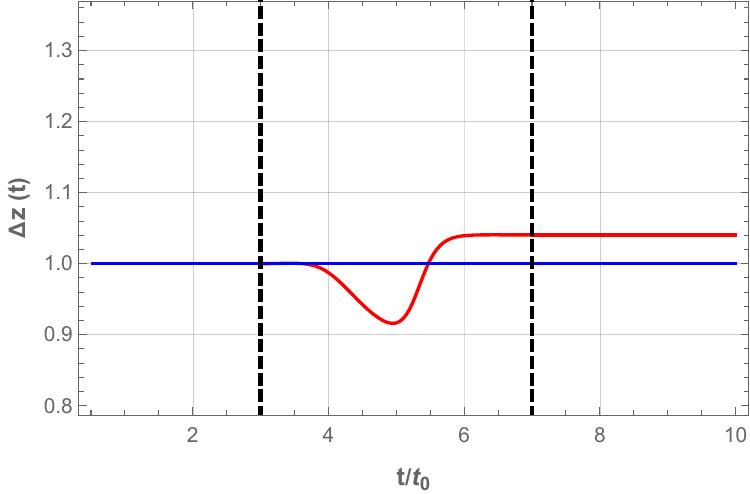}
    \caption{The relative separation between two neighbouring geodesic $\Delta x, \Delta y$ and $\Delta z$ are plotted as function of cosmic time $t/t_{0}$. Blue and Red curves denote the same cases as defined earlier.} 
    \label{fig:2}
\end{figure}

{\bf{GWME in Cartesian coordinates:}} 
The above equations Eq (\ref{rer}) and Eq (\ref{22}) are our first set of major results from which we can directly exhibit GWME as depicted in Figure (\ref{fig:1}) along x-axis, the GW propagation direction and along the transverse y and z-directions. Generally one prefers to restrict the GW sector in a limited time interval so that its effect in generating GWME can be compared with time zones where GW is absent. Strictly speaking, in our linearized framework, this could have been achieved by constructing a wave packet out of the GW modes of Eq (\ref{33}). In the present work we have simply introduced a Gaussian damping factor that restricts the GW in a limited time zone. The relative separations between two neighbouring geodesics, $\Delta x_i$, are plotted in the Figure (\ref{fig:2}) where the blue and red curves represent the absence and presence of GW and subsequently GWME respectively. The displacement GWME is directly visible in the relative separation of neighbouring geodesics before and after the passing of the GW.

\begin{figure}[H]
    \centering
    \includegraphics[width=0.32\linewidth]{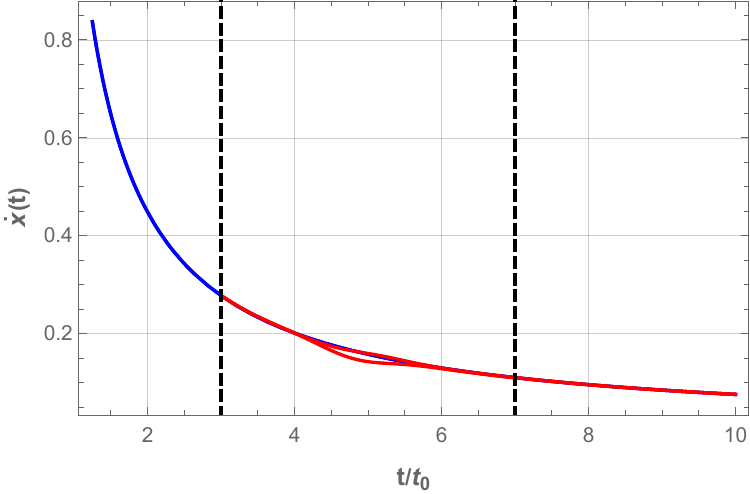}
    \includegraphics[width=0.32\linewidth]{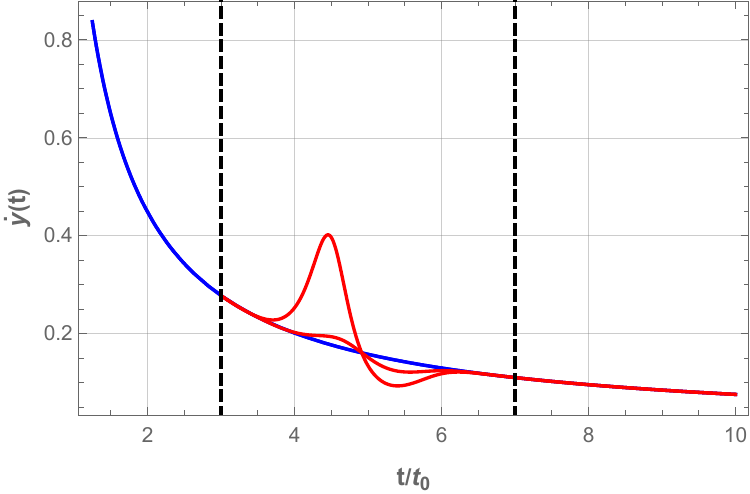}
    \includegraphics[width=0.32\linewidth]{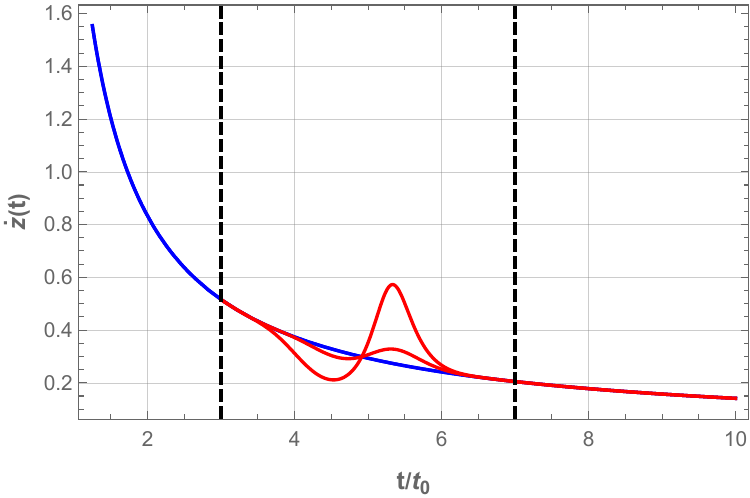}
    \caption{The $\dot{x}, \dot{y}$ and $\dot{z}$-coordinate of velocity geodesic are plotted as a function of cosmic time $t/t_{0}$. Blue curve denotes the results in absence of GW profile and Red curve denotes the same in  presence of GW profile.}
    \label{fig:3}
\end{figure}

\begin{figure}[H]
    \centering
    \includegraphics[width=0.32\linewidth]{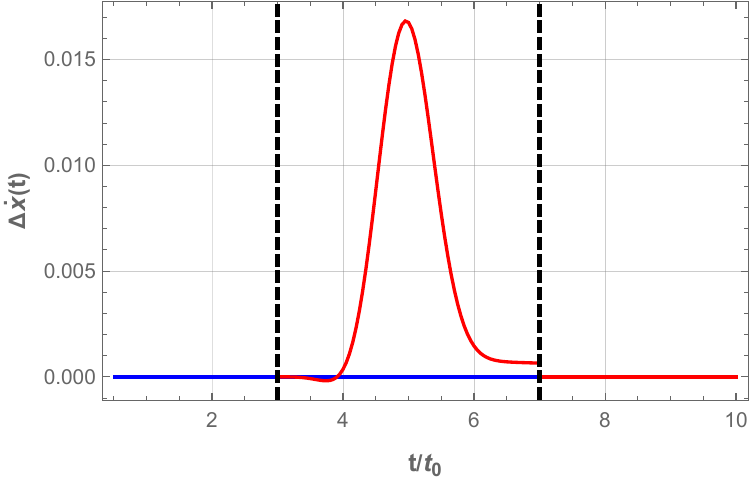}
    \includegraphics[width=0.32\linewidth]{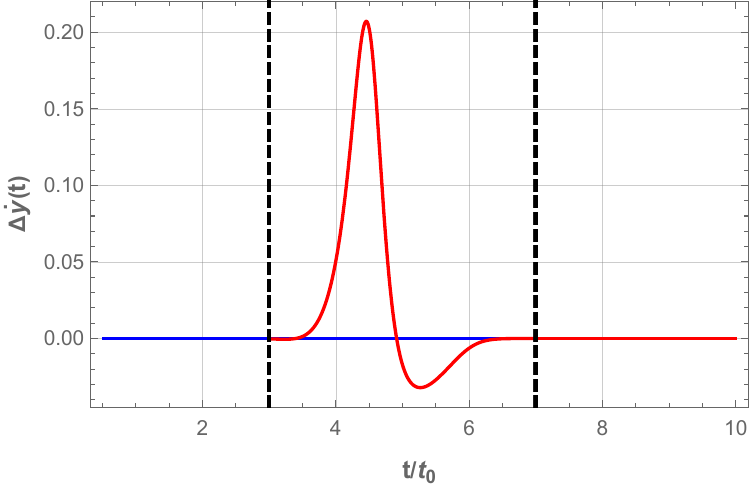}
    \includegraphics[width=0.32\linewidth]{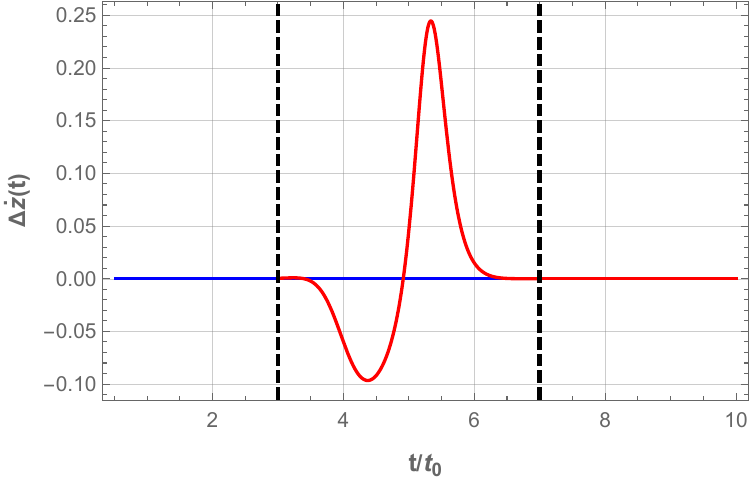}
    \caption{The relative separation between two neighbouring velocity geodesic $\Delta \dot{x}, \Delta \dot{y}$ and $\Delta \dot{z}$ are plotted as function of cosmic time $t/t_{0}$. Blue and Red curves denote the same cases as defined earlier.}
    \label{fig:4}
\end{figure}

In Figure (\ref{fig:3}) individual velocities $\dot x,\, \dot y\, ,\dot z$ are plotted for two geodesics and whereas  the differences $\Delta\dot x,\, \Delta\dot y\, ,\Delta\dot z$ are shown in  Figure (\ref{fig:4}). Interestingly, there is no significant velocity GWME in any of the profiles, which is consistent with Figure (\ref{fig:2}) where the geodesic separations remain constant.

\begin{figure}[H]
    \centering
    \includegraphics[width=0.4\linewidth]{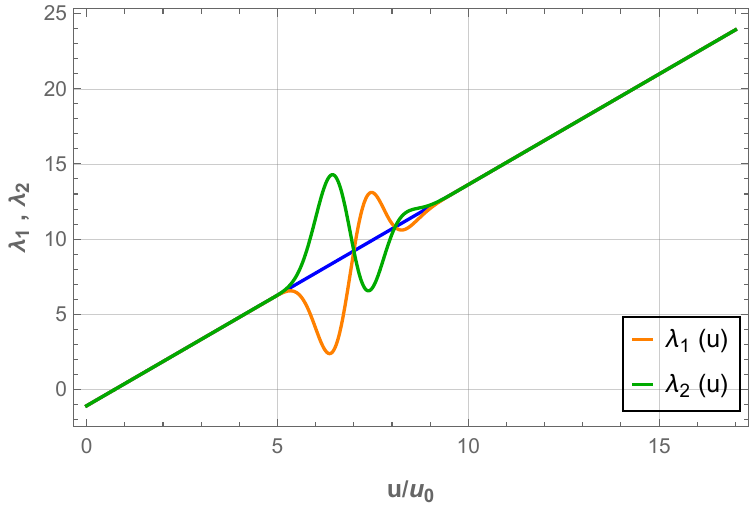}
    \includegraphics[width=0.4\linewidth]{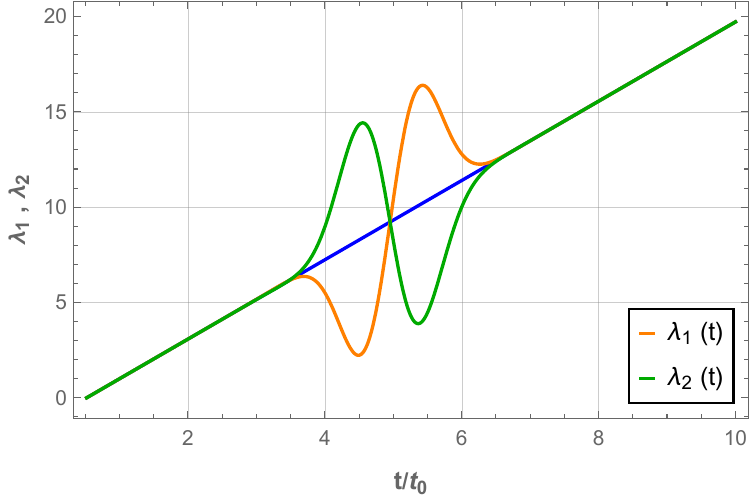}
    \caption{Plotting of $\lambda_1=(a^{2}+\sqrt{2}h)$ (orange curve) and $\lambda_2=(a^{2}-\sqrt{2}h)$ (green curve) as a function of the retarded null coordinate $u$ (in the left figure) and as a function of cosmic time $t$ (in the right figure).}
    \label{fig:9}
\end{figure}

{\bf{GWME in Brinkmann coordinates:}} The generic Brinkmann form for a Plane Wave spacetime metric is given by 
\begin{equation}
ds^2=-2du~dv+H(u,y,z)du^2 +dy^2+dz^2
    \label{b1}
\end{equation}
where $u,v = (t\mp x)/\sqrt{2}$ are the light-cone coordinates. Notice that the function $H(u,y,z)$, carrying the GW features, is independent of $v$ and is widely used in  Plane GW spacetimes \cite{kar} that {\it{assumes}} a $v$-independent $H$-profile. This allows for a plane GW propagating along x-direction, with $y$ and $z$ being transverse coordinates, resulting in a simplified  geodesic equations yielding analytic results. Recall that we have been critical about the physicality of these plane wave spacetimes and in fact our choice of GW profile, as given by Eq (\ref{33}) is not of a plane wave form, since in our case $H(x,t)=H(u,v)$ depends on $v$ as well as $u$. 
First of all, we rewrite the line element, given by Eq (\ref{metric_GW}) in terms of $u,v$ coordinates as 
\begin{equation}
    \label{1metric_GW}
    ds^{2} \, = \, \frac{a^2-1}{2}(du^2 +dv^2) -2 \, \frac{a^2+1}{2} \, dudv  + \{a(t)^{2} + h_{0}(t,x) \} \, dy^{2} + \{a(t)^{2} - h_{0}(t,x) \} \, dz^{2} + 2h_{1}(t,x) \, dydz \hspace{0.08cm} . 
\end{equation}
As done earlier, after diagonalizing the transverse y-z sector Eq (\ref{1metric_GW}), the line element becomes
\begin{eqnarray}
    ds^{2} &=& \left({du}^2+{dv}^2\right) H(u,v)+2 {du} {dv}~ G(u,v)+{dy}^2 {\lambda_1}(u,v)+{dz}^2 {\lambda_2}(u,v) \, . 
\end{eqnarray}
In the Brinkmann coordinate, the metric components are given as (remember that $h_{0}(u, v) = h_{1}(u, v)$)
\begin{eqnarray}
    H(\text{u}, v) &=& \frac{1}{2} \left\{a(u,v)^2 - 1 \right\} \hspace{1.2 cm} , \hspace{0.4 cm} G(\text{u}, v) \hspace{0.1 cm} = \hspace{0.1 cm} \frac{1}{2} \left\{a(u,v)^2 + 1 \right\} \, , \nonumber \\   
    \lambda_1(\text{u},v) &=& \{a(u,v)^2+\sqrt{2} h(u,v) \} \hspace{0.4 cm} , \hspace{0.4 cm} \lambda_2(\text{u},v) \hspace{0.1 cm} = \hspace{0.1 cm} \{a(u,v)^2-\sqrt{2} h(u,v) \} \, . 
\end{eqnarray}
Exploiting the reparametrization invariance, we fix the gauge $u = \tau$ and obtain the geodesic equations for $v$ as, ($\dot v=dv/du$),
\begin{eqnarray}\label{ll2}
    \hspace{-0.3 cm} \ddot v &+& \frac{\dot v^2}{2 \left(H^2-G^2\right)} \left\{G \left(\frac{\partial H}{\partial u}-2 \frac{\partial G}{\partial v}\right) + H \frac{\partial H}{\partial v}\right\} + \frac{\dot v}{G^2-H^2} \left\{G \frac{\partial H}{\partial v}-H \frac{\partial H}{\partial u}\right\} \nonumber \\
    &+& \, \frac{\dot y^2}{2 \left(G^2-H^2\right)} \left\{H \frac{\partial \lambda_1}{\partial v}-G \frac{\partial \lambda_1}{\partial u}\right\} \, + \, \frac{\dot z^2}{2 \left(G^2-H^2\right)} \left\{H \frac{\partial \lambda_2}{\partial v}-G \frac{\partial \lambda_2}{\partial u}\right\} \, \nonumber \\ 
    &+& \frac{1}{2 \left(G^2-H^2\right)} \left\{H \left(\frac{\partial H}{\partial v}-2 \frac{\partial G}{\partial u}\right)+G \frac{\partial H}{\partial u} \right\} \, = \, 0 \, .  \hspace{-1.6 cm}
\end{eqnarray}
The $y$ and $z$-coordinates still remain the cyclic coordinates, resulting in  first order differential equations
\begin{eqnarray}
    \frac{dy}{du} &=& \frac{(p_{y})_{0}}{2\lambda_1} \hspace{0.7 cm} , \hspace{0.7 cm} \frac{dz}{du} \hspace{0.2 cm} = \hspace{0.2 cm} \frac{(p_{z})_{0}}{2\lambda_2} .  
    \label{ll1} 
\end{eqnarray}
In Figure (\ref{fig:9}) $\{\lambda_1(t, x=0),\, \lambda_2(t, x=0) \}$ are plotted as function of Cartesian time $t$ (in left figure) and $\{\lambda_1(u, v=0),\, \lambda_2(u, v=0) \}$ are plotted as function of Brinkmann null coordinate $u$ (in right figure), showing their opposite behaviour in the GW zone. 

\begin{figure}[H]
    \centering
    \includegraphics[width=0.32\linewidth]{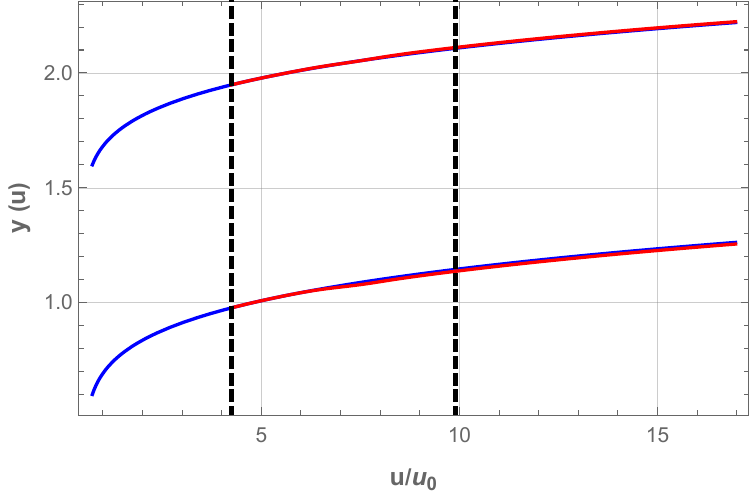}
    \includegraphics[width=0.32\linewidth]{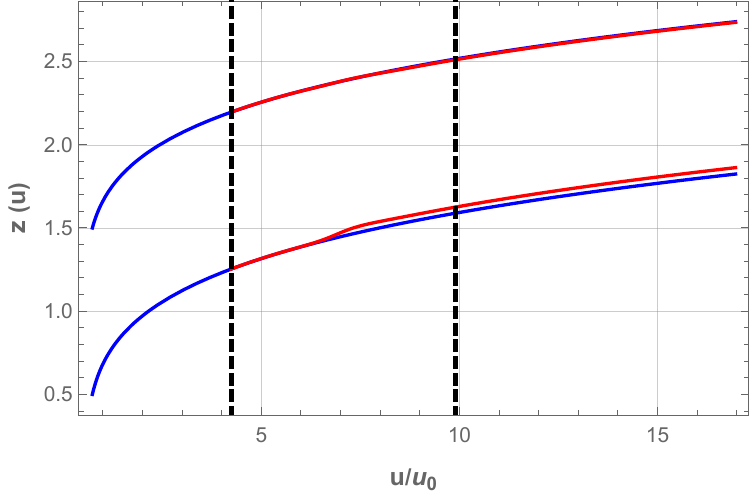}
    \includegraphics[width=0.32\linewidth]{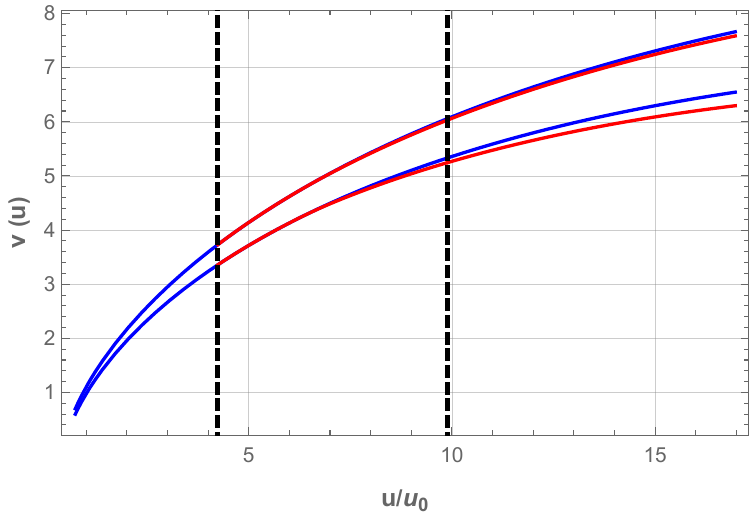}
    \caption{The y, z and v-coordinate of geodesic equation are plotted as a function of null coordinate $u/u_{0}$. Blue curve denotes the results in absence of GW profile and Red curve denotes the same in  presence of GW profile.}
    \label{fig:5}
\end{figure}

\begin{figure}[H]
    \centering
    \includegraphics[width=0.32\linewidth]{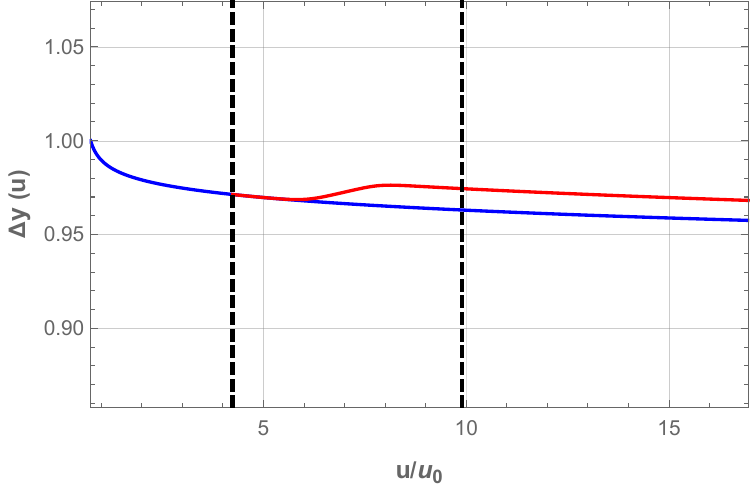}
    \includegraphics[width=0.32\linewidth]{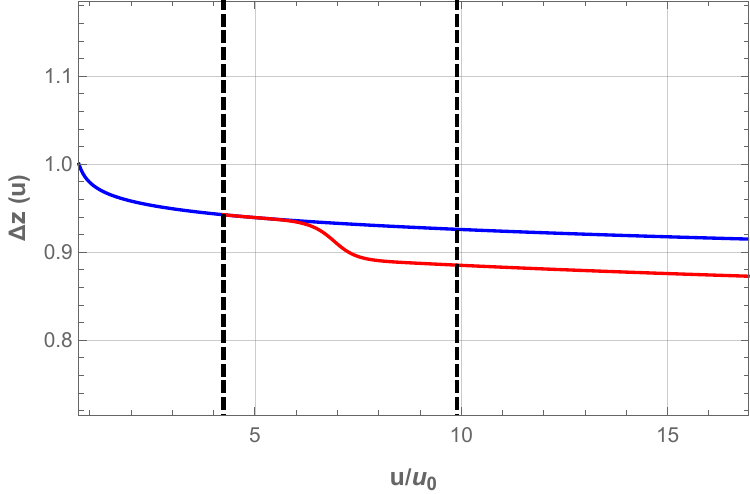}
    \includegraphics[width=0.32\linewidth]{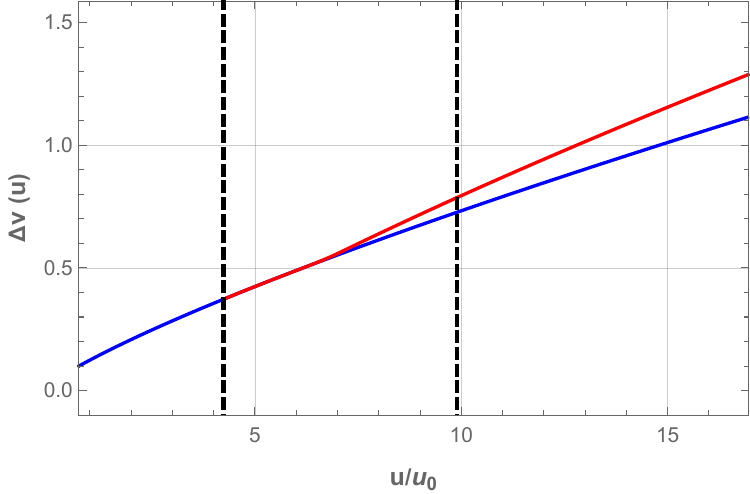}
    \caption{The relative separation between two neighbouring geodesic $\Delta y, \Delta z$ and $\Delta v$ are plotted as function of null coordinate $u/u_{0}$. Blue and Red curves denote the same cases as defined earlier.}
    \label{fig:6}
\end{figure}

\begin{figure}[H]
    \centering
    \includegraphics[width=0.32\linewidth]{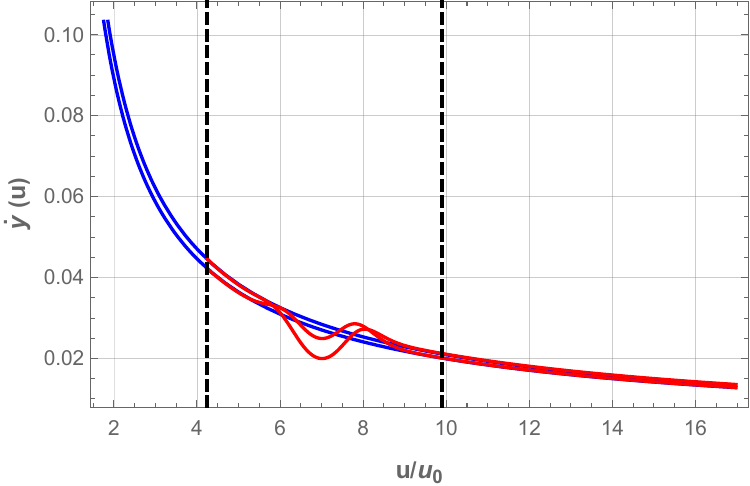}
    \includegraphics[width=0.32\linewidth]{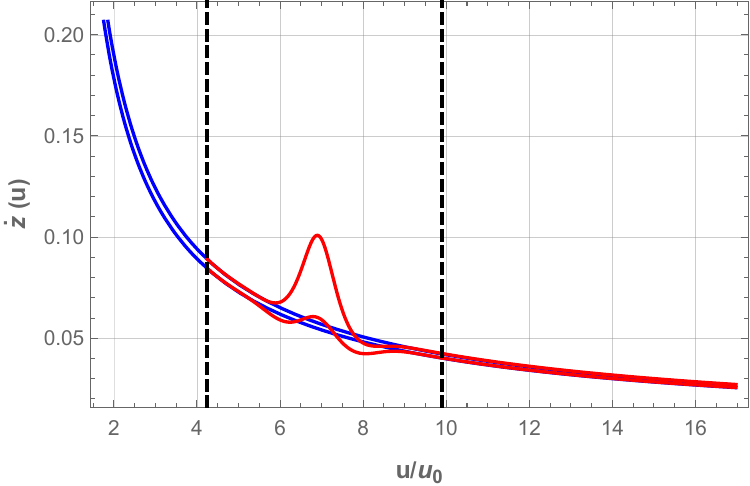}
    \includegraphics[width=0.32\linewidth]{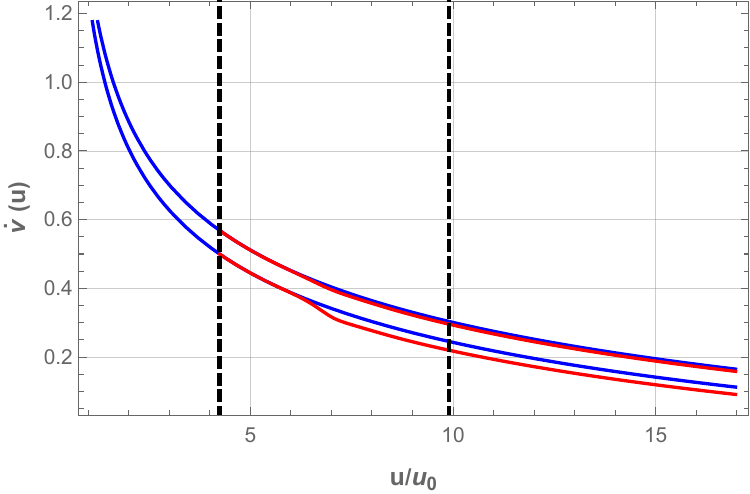}
    \caption{The $\dot{y}, \dot{z}$ and $\dot{v}$-coordinate of velocity geodesic are plotted as a function of null coordinate $u/u_{0}$. Blue curve denotes the results in absence of GW profile and Red curve denotes the same in  presence of GW profile.}
    \label{fig:7}
\end{figure}

\begin{figure}[H]
    \centering
    \includegraphics[width=0.32\linewidth]{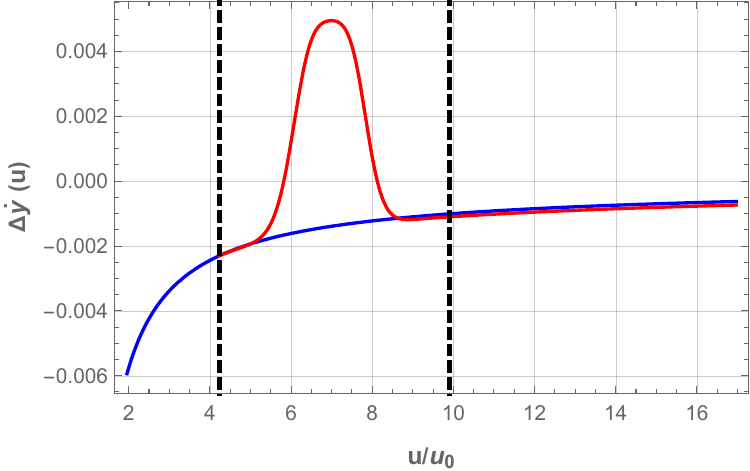}
    \includegraphics[width=0.32\linewidth]{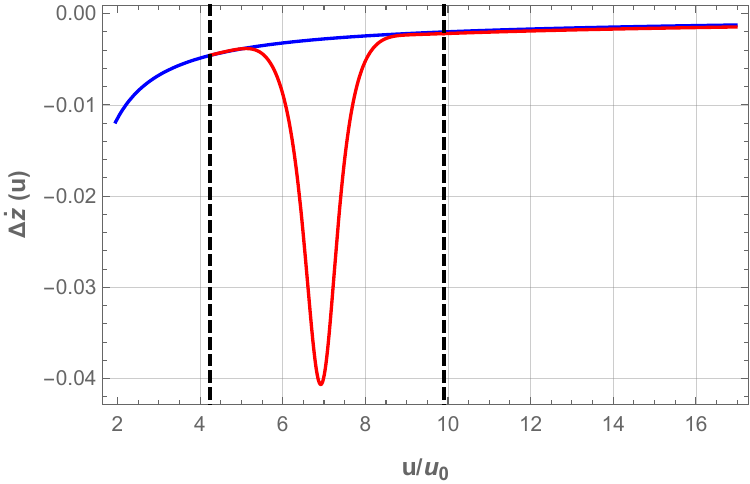}
    \includegraphics[width=0.32\linewidth]{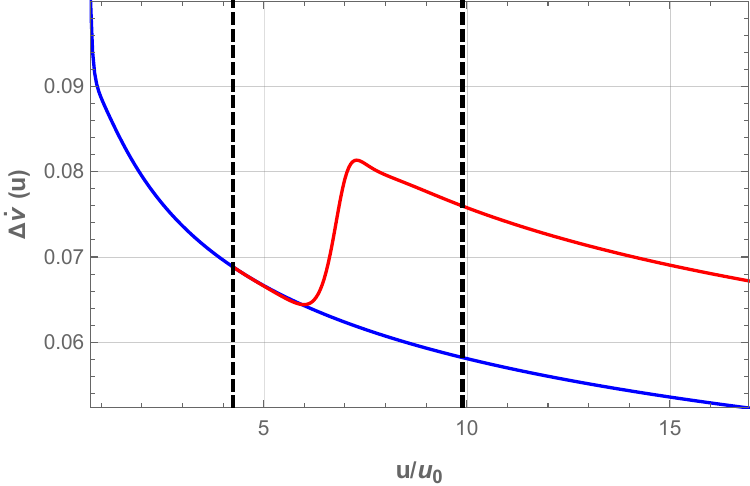}
    \caption{The relative separation between two neighbouring geodesic $\Delta \dot{y}, \Delta \dot{z}$ and $\Delta \dot{v}$ are plotted as function of null coordinate $u/u_{0}$. Blue and Red curves denote the same cases as defined earlier.}
    \label{fig:8}
\end{figure}

We re-establish the presence of GWME in Brinkmann setup, taking recourse to numerical computation with the geodesic profiles as shown in Figure (\ref{fig:5}) for the transverse $y(u), z(u)$ and longitudinal $v(u)$-directions respectively. It is worthwhile to observe that, in the transverse sector, the GWME is more pronounced for $\Delta z(u)$ rather than $\Delta y(u)$. The probable reason can be understood from Figure (\ref{fig:9}) where $\lambda_1$ (orange curve), $\lambda_2$ (green curve) are plotted against $u/u_0$. Notice that $\lambda_2$ comes very close to zero and hence, from the Eq (\ref{ll1}), we found that its impact on $\dot{z}$ is much larger than the effect of $\lambda_{1}$ on $\dot{y}$. We stress that in the numerical results, as shown in the Figures (\ref{fig:5}, \ref{fig:6}), the $v$-dependence of $y(u,v)$ and $z(u,v)$-geodesic is incorporated by first substituting $\dot y, \dot z$ from  (\ref{ll1}) to the Eq (\ref{ll2}) so that $v(u)$-geodesic becomes completely decoupled from the transverse y and z-coordinate. Substituting the solution of  $v(u)$ further in the  (\ref{ll1}) to compute $y(u)$ and $z(u)$. Obviously our GWME is more general in comparison with existing works since in the latter, unphysical examples of GW profile are chosen that are independent of $v$. The Displacement GWME is clearly visible in the transverse $y,z$ and longitudinal $v$ coordinate through the relative separation of neighbouring geodesics before and after the passage of GW. A somewhat surprising feature is that for $v$ coordinate, $\Delta v$ continues to increase in the later time zone where the GW ceases to exist (see \cite{kar} for similar behaviour). In Figures (\ref{fig:7}) and (\ref{fig:8}) we plot the velocities in null coordinate $u$ showing that a non-zero GWME exists in the case of $v$-velocity. We emphasize that our analysis in Brinkmann coordinate keeps $v$-contribution throughout so that for the first time this type of general form of GWME is being reported here.

{\bf{Approximate analytic forms for transverse coordinates:}} We impose a simplifying assumption of small $v$ and put $v=0$. It should be kept in mind that $v(u)$ being a dynamical variable, fixing $v$ can give rise to complications. Strictly speaking $v=0$ should be treated as a constraint leading to a Hamiltonian constraint analysis a la Dirac {\cite{dir}. In our restricted framework, we sidestep this issue by keeping the option of a dynamical $v(u)$ open and study the $v$-geodesic equation as well, albeit on $v\approx0$ hypersurface. \\ 

Next we choose $v=0$ gauge where the induced metric, with $t=u/\sqrt{2},~x=-u/\sqrt{2}$, is given by 
\begin{equation}
    ds^{2} \, = \, \frac{a^2-1}{2} \, du^2   + \lambda_1(u) dy^{2} + \lambda_2(u) dz^{2} \hspace{0.4 cm} ; \hspace{0.4 cm} \lambda_{1,2} \, = \, a^{2}(u) \pm \sqrt{h_0^2(u)+h_1^2(u)} \hspace{0.1 cm} = \hspace{0.1 cm} a^{2}(u) \pm \sqrt{2}h(u) \, . 
    \label{xx1}
\end{equation}

Considering the metric perturbations $h(u)$ to be small enough, the approximate form of the $y$-geodesic equation can be rewritten as
\begin{eqnarray}
    \frac{dy}{du} \, = \, \frac{\tilde{p}_{y0}}{2 (a^{2} + \sqrt{2}h)} \hspace{0.3 cm} \approx \hspace{0.3 cm} 
    \frac{\tilde{p}_{y0}}{2 a^{2}} \left(1 - \frac{\sqrt{2} h}{a^{2}} \right) . 
\end{eqnarray}
Again considering the arguments of sine and cosine functions (present in the GW profile) to be small enough such that one can retain only the first order terms of their Taylor expansion, yields the following
     \begin{eqnarray}
    \frac{dy}{d \omega}= k_{1} + k_{2} \sqrt{\omega} + \frac{k_3}{\sqrt{\omega}} + \frac{k_4}{\omega^{3/2}}+\frac{k_5}{\omega^{5/2}} + \frac{k_6}{\omega} + \frac{k_7}{\omega^2} \hspace{0.1 cm} ,   
\end{eqnarray}
where $\omega = \sqrt{41 \Omega_{\lambda}}/2 \left\{1 + 2H_{0}t_{0} \left(\frac{u}{\sqrt{2}} - 1 \right) \right\}$. An approximate analytic form of $y(\omega)$ is given by
\begin{eqnarray}
    y(\omega) &=& c_{1} + k_{1} \omega + \frac{2k_{2}}{3} \omega^{3/2} - \frac{2 k_5}{3 \omega^{3/2}} + 2k_{3} \sqrt{\omega} - \frac{2 k_4}{\sqrt{\omega}} - \frac{k_7}{\omega} + k_6 \log (\omega) \hspace{0.1 cm} . 
\end{eqnarray}
The coefficients $k_i, \, i=1..7$, expressed in terms cosmological density parameters $(\Omega_{\lambda}, \Omega_{k}, \Omega_{r})$, constants of motion $(\tilde{p}_{y0}, \tilde{p}_{z0})$ and present day value of cosmological parameters $(a_{0}, H_{0}, t_{0})$ are given the Appendix C. An analogous form for $z(\omega)$ is available. The comparison depicted in Figure (\ref{fig:10}) shows that  the approximate analytic mimics the exact one reasonably well.

\begin{figure}[H]
    \centering
    \includegraphics[width=0.5\linewidth]{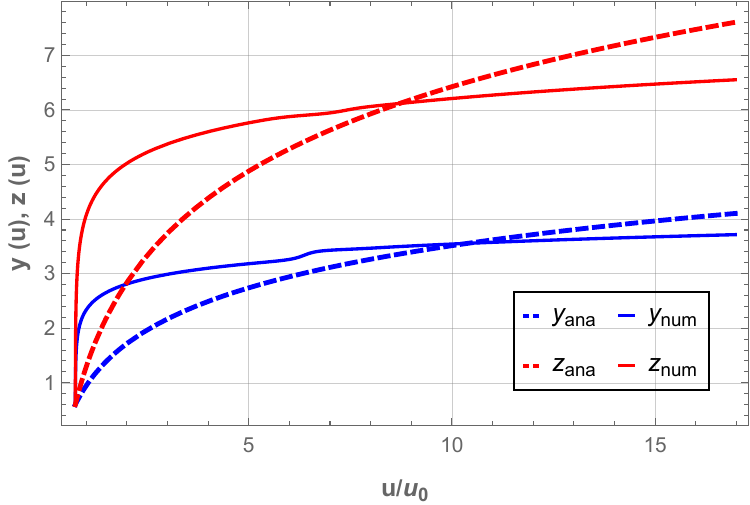}
    \caption{Numerical (solid lines, Blue/Red) and Approximate (dotted lines, Blue/Red) solutions of the y and z-geodesic equations are plotted in terms of Brinkmann null coordinate $u/u_{0}$.}
    \label{fig:10}
\end{figure}

{\bf{Highlights of our work:}} \\ To summarize, the present work consists of 
\begin{itemize}
    \item Establishing convincingly the presence of GWME in a new domain: the TC condensate in $(R + \alpha R^2)$-Starobinsky model.
    \item Instead of an idealized and physically unreasonable plane wave structure, we have used the GW structure generated directly in the TC condensate picture.
    \item Results in both Cartesian and Brinkmann coordinates are presented. This is the first time that results using GWME with more general coordinate dependence, (i.e. depending on $x,t$ or $u,v$) are revealed. 
    \item It might be interesting to suggest some back of the envelope numbers concerning the displacement Memory Effect. For reasonable values of the cosmological parameters, taking initial separations between the two neighbouring geodesics in question as $\Delta x=\Delta y=\Delta z=1.0$ (in appropriate units), the final separations (after the passage of GW) are found out to be $\Delta x = 1.0589 \, , \, \Delta y = 1.1628 \, , \, \Delta z = 1.0405 $ revealing GWME in Cartesian coordinates. However the situation is slightly more complicated in Brinkmann coordinates since $y(u),z(u)$ and $v(u)$ change  both in presence and absence of GW, albeit with different rates, thus leading to GWME (compare  Figures (\ref{fig:2}) and (\ref{fig:6}) for Cartesian and Brinkmann coordinates respectively). For this case, with $\Delta y=\Delta z=\Delta v=1.0$ at $u=0$ one obtains, at $u = 16.69$, in absence of GW throughout, $\Delta y = 0.9577 \, , \, \Delta z = 0.9154 \, , \, \Delta v = 1.1005  $, whereas 
    for a passage of GW,     $\Delta y_{GW} = 0.9683 \, , \, \Delta z_{GW} = 0.8732 \, , \, \Delta v_{GW} = 1.2709  $ (in same appropriate unit), once again showing non-trivial GWME. The only significant velocity GWME is shown through the difference of v-coordinate as depicted in Figure (\ref{fig:8}), more precisely $ \Delta \dot{v} = 0.0524  $ in absence of GW and $ \Delta \dot{v}_{GW} = 0.0674  $ after passage of GW.
\end{itemize}
As a future problem we would like to consider the non-linear GWME in our TC condensate scenario.\\

{\bf{Acknowledgement:}} It is a pleasure to thank Professor Sayan Kar for valuable discussions. AM would like to thank Sourav Pal for the helpful discussions regarding  Mathematica.

\end{document}